 \documentclass{optica-article}
 \journal{opticajournal} % for journals or Optica Open
 \articletype{Research Article}
\usepackage{lineno}

\begin{document}

\title{Multiplexed Processing of Quantum Information Across an Ultra-wide Optical Bandwidth}

\author{Alon Eldan, Ofek Gillon, Asher Lagemi, Elai Fishman Furman, Avi Pe'er,\authormark{*}}

\address{Department of Physics and QUEST Center for Quantum Science and Technology, Bar-Ilan University, Ramat Gan 5290002, Israel}

\email{\authormark{*}avi.peer@biu.ac.il}

\begin{abstract}
Protocols for processing of quantum information are the foundation of quantum technology, enabling to share secrets at a distance, teleport quantum states, and to implement quantum computation. While many protocols were realized, and even commercialized, the throughput and processing speed of current protocols is limited by the narrow electronic bandwidth of standard measurement devices (typically in the MHz-to-GHz range), which is orders-of-magnitude lower than the optical bandwidth of available quantum optical sources (10-100 THz), indicating that the bandwidth resource is dramatically underutilized in current quantum optical technology. We present a general concept of frequency multiplexed quantum channels and a set of methods to process quantum information efficiently across the available optical bandwidth. Using a broadband source of squeezed light, spectral manipulation methods and parametric homodyne detection, we are able to generate, process and measure all the channels in parallel, thereby harnessing the optical bandwidth for quantum information in an efficient manner. We exemplify the concept through two basic protocols: Multiplexed Continuous-Variable Quantum Key Distribution (CV-QKD) and multiplexed continuous-variable quantum teleportation. The multiplexed QKD protocol is demonstrated in a proof-of-principle experiment, where we successfully carry out QKD over 23 uncorrelated spectral channels, with capability to detect eavesdropping in any channel. These multiplexed methods (and similar) will enable to carry out quantum processing in parallel over hundreds of channels, potentially increasing the throughput of quantum protocols by orders of magnitude.
\end{abstract}
%\maketitle

% XXXXXXXXXXXXXXXXXXXXXXXXXXXXXXXXXXXXXXXXXXXXXXXXXXXXXXXXX %
\section{Introduction}
\label{section:main}
In the decades since the conception of quantum information in the 1980s, many practical applications were developed, ranging form secure communications \cite{BENNETT20147, BBM92Implementation, Eckert91} and transmission of quantum information \cite{QuantumTeleportation}, through quantum sensing schemes \cite{QuantumInformationCaves, QuantumPlasmonicSensing, EntanglementSensing, DistributedQuantumSensing} to quantum computation \cite{QuantumTuringMachine, GoogleQuantumSupremacy, TrappedIons, NeutralAtomsQuantumComputing}. All these applications rely on some uniquely quantum property, such as entanglement, squeezing, etc., to encode, process and decode the desired quantum information. Practically any degree of freedom of light can carry quantum information, in either discrete photons  or in continuous variables, such as polarization \cite{BENNETT20147, BBM92Implementation}, spatial mode \cite{SpatialModeQKD}, field quadratures \cite{QuantumPlasmonicSensing, SqueezedStateQKD}, frequency \cite{FrequencyCodedQKD} and time \cite{TimeDivisionQKD}.
%Or material degrees of freedom, such as the energy levels of trapped ions \cite{TrappedIons} and neutral atoms \cite{NeutralAtomsQuantumComputing} or the phase / charge of Josephson junctions \cite{JosephsonCharge, JosephsonCharge2, JosephsonPhase}.

Specifically, the quadratures of the optical electric field are the backbone of many protocols of optical quantum processing. Classically, the quadratures $x,y$ are the cosine and sine components of the optical field at frequency $\omega$: $E(t)=x \cos\left(\omega t\right) + y \sin\left(\omega t\right) = \left|a\right| \cos\left(\omega t+\varphi\right)$, where $a=\left|a\right|e^{i\varphi}$ is the complex field amplitude. The quadratures are the real $x=a+a^*$ and imaginary $y=i\left(a-a^*\right)$ components of the complex amplitude. Quantum mechanically, the optical quadratures are analogous to the canonical position and momentum, defined as $x=\left(a+a^\dagger\right)/2,y=i\left(a-a^\dagger\right)/2$ with the canonical commutation relation $[x,y]=i/2$ and uncertainty relation $\Delta x \Delta y \geq 1$. One can therefore encode, store, process and decode quantum information on the optical quadratures. Specifically, squeezed states of light, where the fluctuations of one quadrature are squeezed below the vacuum level, while the orthogonal quadrature is stretched, can be utilized for quantum processing and represent quantum-bits (qubits) at both low and high squeezing levels, as we show hereon.

The focus of this paper is to harness the bandwidth of ultra-broadband sources of squeezed light to drastically enhance the rate of quantum optical processing. Most generally, quantum information processing can be broken into three primary stages: generation of the quantum state, manipulation of the state, and its measurement. While the speed of each stage can be limited by different factors, the primary bottleneck in quantum optical protocols is the measurement, where the relatively slow response of photo-detectors limits the processing rates at several orders-of-magnitudes below the optical bandwidth of available sources, even with the fastest available detectors. In particular, sources of broadband squeezed light with 10-100THz of bandwidth (up to an optical octave) are readily available \cite{PulseBroadbandSource, OctaveSqueezedLight}, as well as methods of broadband manipulation using pulse shaping in the spectral domain \cite{BroadbandSpectralShaping, UltrabroadbandPulseShaping}. In contrast, the bandwidth of traditional measurement techniques was always limited by the narrowband electronic response of optical detectors, in the MHz-to-GHz range. 

\begin{figure}[ht!]
\centering\includegraphics[width=\textwidth]{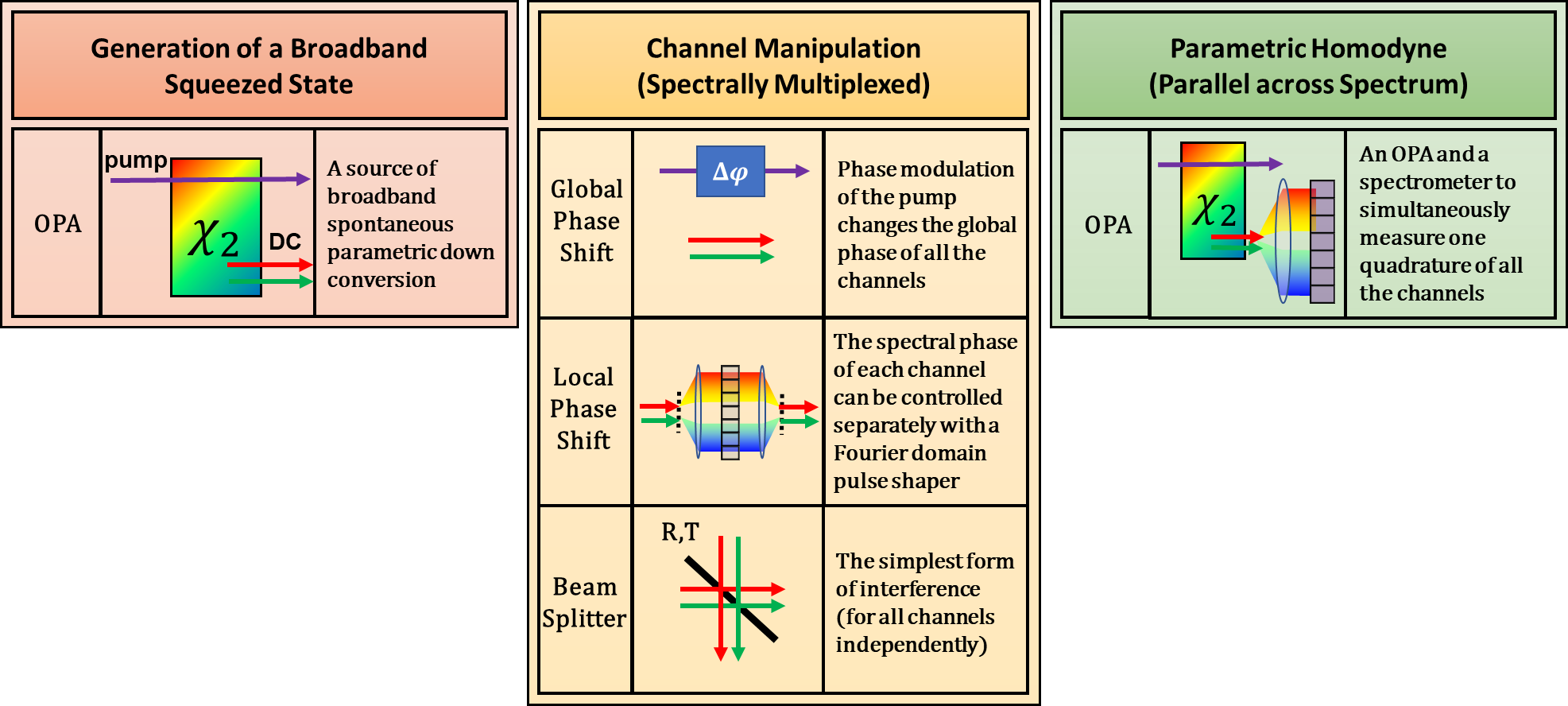}
\caption{The tool-set for multiplexed processing with broadband squeezed light, decomposed to generation, manipulation and detection of quantum states in a multiplexed manner. State generation employs a broadband optical parametric amplifier (OPA), which generates a broad spectrum of two-mode squeezed light from a narrowband pump, either by spontaneous parametric down-conversion (SPDC) or four-waves mixing (FWM). State manipulation is performed by spectrally-resolved phase shifters and by beam splitters, which allow to control the quadratures of the quantum states separately for each channel. Finally, the detection employs parametric homodyne in an additional OPA, where the chosen quadrature is amplified, followed by a spectrometer for simultaneous spectrally resolved measurement of that quadrature. Later we employ this tool-set  to assemble two quantum protocols (multiplexed QKD and quantum teleportation), as two major examples. }
\label{fig:general_building_blocks}
\end{figure}

Luckily, this electronic bandwidth limit was recently overcome with multimode SU1,1 interferometry and with parametric homodyne detection \cite{OpticalParametricHomodyne}, which enable to measure an optical quadrature of interest across a wide, practically unlimited optical spectrum, opening a path to much faster quantum processing. We present a general approach for parallel processing of quantum information, encoded across the entire optical spectrum of the quadratures of broadband two mode squeezed light. We highlight a set of tools,  illustrated in figure \ref{fig:general_building_blocks}, to simultaneously generate, manipulate and measure quantum information over multiple frequency channels, up to $10^3 - 10^4$ channels in realistic configurations, limited only by the available optical bandwidth. 

Here, the tools of figure \ref{fig:general_building_blocks}  helped us to develop a multiplexed version for two of the most important quantum processing protocols:  
\begin{enumerate}
    \item \textit{Multiplexed quantum teleportation};  Teleportation is a critical stepping stone of quantum computation in particular and of quantum technologies in general. For example, in measurement-based quantum computation, which is the chosen scheme for leading companies, such as PsiQuantum \cite{PsiQuantum} and Xanadu \cite{Xanadu}, quantum teleportation is key for implementing all 2-qubit gates \cite{DeterministicGateTeleportation, QuantumTeleportationComputationalPrimitive}. Thus, multiplexed quantum teleportation, as we present hereon can significantly scale-up the number of qubit-gates and speed up the processing accordingly.
    \item \textit{Multiplexed QKD} (BB84-like); QKD provides safe key sharing between two distant parties, whose security is physically-proven. Although many QKD protocols were already demonstrated, the communication speed in all of them is limited by the detection bandwidth of a single channel, which is bound by electronic performance. Our QKD protocol can bridge the orders of magnitude gap between the wide bandwidth of light sources and the narrow bandwidth of detectors. To highlight the capabilities of our set of tools, we demonstrate our multiplexed QKD in a proof-of-principle experiment, which successfully processed information over 23 parallel spectral channels.
\end{enumerate}
And yet, those two protocols serve primarily to exemplify the general utility of the presented toolkit - to multiplex quantum protocols with entangled photons / squeezed light. In general, we expect our set of tools to be useful also for many other quantum processing technologies thereby enhancing their processing throughput by orders of magnitude. 

\section{Multiplexed QKD}
\label{section:QKD_overview}

\begin{figure}[ht!]
\centering\includegraphics[width=14cm]{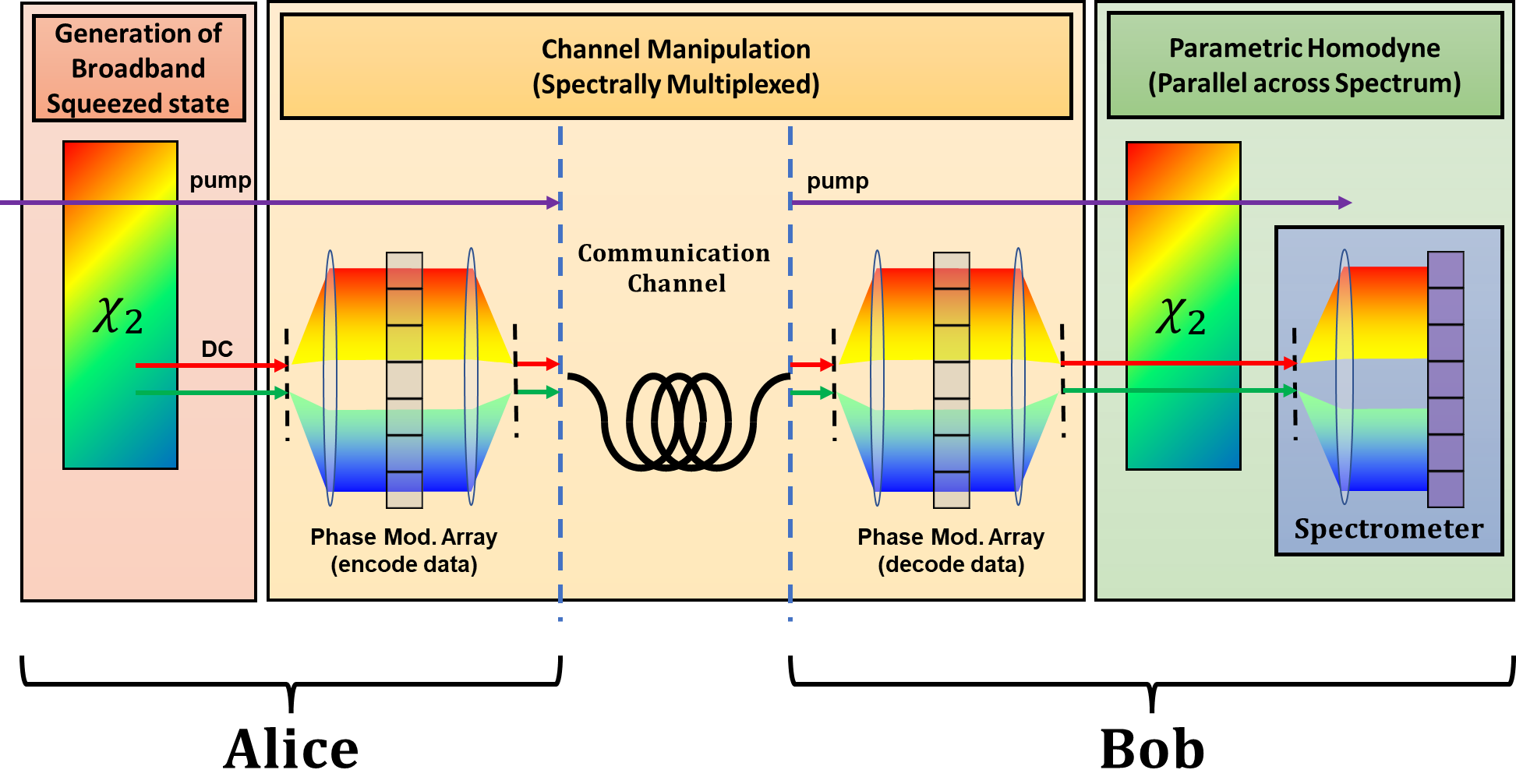}
\caption{\textbf{The multiplexed QKD protocol}. To encode the data Alice employs a source of broadband entangled photon pairs (two-mode, weakly squeezed light \cite{TwoModeQuadrature}), generated using an OPA with broadband phase-matching. Alice modulates the spectral phase of the squeezed light using a Fourier-domain spectral shaper that consists of a dispersive element, such as grating or prism to spatially separate the frequencies and a spatial light modulator (SLM) to vary the phase of each spectral channel independently. The modulated light is then sent to Bob via an optical fiber along with the pump that serves as a local oscillator for all the frequencies. To decode Alice's information, Bob chooses a measurement basis for each channel by modulating the spectral phase of the squeezed light with his spectral shaper (identical to Alice's) and passes the light through a second OPA (completing the SU(1,1) interferometer). Finally, Bob measures the spectral intensity with a spectrometer trying to detect single photons in each channel.}
\label{fig:QKD_scheme}
\end{figure}

As a first example for a multiplexed protocol, let us present (and later demonstrate experimentally) the simple protocol of multiplexed QKD with entangled bi-photons, illustrated in figure \ref{fig:QKD_scheme}. Our scheme forms a continuous-variable analog of the BB84 protocol \cite{BENNETT20147, LongDistanceCVQKD, HighRateCVQKD} using an unseeded SU(1,1) interferometer \cite{ChekhovaOu}, where two OPAs are set in series and pumped by a single pump. The phase-sensitive OPAs amplify / attenuate one quadrature of light with \textbf{no added noise} \cite{OpticalParametricHomodyne, ChekhovaOu}, indicating that the generation of bi-photons in the two OPAs can fully interfere for any gain settings, depending on the phase of the signal-idler pairs relative to the pump in the second OPA. The pump laser therefore serves as the local oscillator to decode the phase of each signal-idler pair. 

In our protocol, Alice holds the first OPA of the interferometer and Bob holds the second. Alice's OPA is unseeded (vacuum input) and operates at low gain, generating randomly time-energy entangled bi-photons through either spontaneous parametric down-conversion (SPDC) or spontaneous four-waves mixing (SFWM). Alice encodes information on the broad spectrum of bi-photons from her OPA using a Fourier-domain spectral shaper that modulates the spectral phase of the light using a spatial phase modulator. Bob can then complete the interferometer with his OPA, thereby allowing him to measure the encoded phase (see figure\ref{fig:QKD_scheme} and caption for details). Since an unseeded OPA with a broad phase-matching generates a wide spectrum of signal-idler pairs, the different frequencies within this spectrum can be used as separate QKD channels.

This leads us to the following 4-steps protocol with the configuration of  figure \ref{fig:QKD_scheme}:
\begin{enumerate}
	\item To encode information, Alice modulates the phase of each signal-idler channel in one of two mutually unbiased bases (chosen at random): Basis 1 uses $\phi=0$ (constructive interference) for logic '1' and $\phi=\pi$ (destructive) for logic '0', whereas basis 2 employs $\phi=\pm\frac{\pi}{2}$. After the modulation of the spectral phase of all the channels in parallel (e.g. using a Fourier-domain pulse shaper) Alice sends the phase modulated spectrum to Bob together with a phase reference (e.g. her pump laser).
	\item To detect the information, Bob randomly chooses a measurement basis (for each channel separately) using another spectral phase modulator - by setting the phase to $0$ for basis 1 or to $\frac{\pi}{2}$ for basis 2 (relative to the phase reference he received from Alice). The, Bob passes the light again through his OPA, where the SU(1,1) interference occurs. Bob measures the spectrally resolved light intensity with a spectrometer, which reflects the number of photons in each channel at the output of the complete SU(1,1) interferometer. If Bob sets the correct basis for a channel, the interference of that channel at the output will be either fully constructive (high probability for photo-detection) or destructive (low probability) and Bob will be able to detect Alice's phase. However, if Bob sets the phase to the wrong basis, his interference will be intermediate for both '0' or '1', preventing Bob from deducing the encoded information.
	\item After the communication is complete, Alice and Bob use a public channel to compare their bases for each channel, keeping only the bits where the encoding and decoding bases matched.
	\item Finally, to detect a possible eavesdropper, Alice and Bob compare a fraction of their data, searching for errors that Eve's measurements could have introduced (just like any other QKD protocol).
\end{enumerate}

This 4-steps protocol formulates the physical layer of a multiplexed QKD communication, i.e. forms a baseline of spectral channels on which QKD can be performed by various protocols of data-exchange. Specifically, a higher protocol should define logical components that are critical for a practical QKD implementation, such as clock synchronization and error-mitigation. For example, the ambiguity between the destructive interference event (bit '0') and the no-photon event (communication fail) can be mitigated if Bob divides his integration time into two time-windows, and flips the measurement phase in the middle by $\pi$ (similar to differential signaling in classical communication). Thus, a successful detection event for Bob must include one photon in exactly one of the time-halves (the logical '0' or '1' is dictated by their order). This allows Bob to identify errors and discard no-photon events. As there are many possible ways to implement such higher-level protocols, we will leave their optimization to future research.

The security of each channel within our scheme can be analyzed similar to the standard analysis of the BB84 protocol, as summarized hereon (the complete derivation of the security is given in supplement 1). Assuming a weak parametric gain in the OPAs, we can employ the perturbative quantum propagator through the nonlinear crystal \cite{Shaked2014}, as $\hat{U}(t) = e^{i\hat{H}t} \approx
1 + i\hat{H}t = 1 + ig_{\omega} \hat{a}^{\dagger}_{\omega} \hat{a}^{\dagger}_{-\omega}$, where $\hat{a}_{\pm \omega}$ represent the field operators of the signal-idler mode pair at $\omega_{p}/2 \pm \omega$ and $g_{\omega}$ is the complex gain of that pair , which includes the interaction time $t$ (proportional to the crystal length) and the relative phase between the pump and the pair. Assuming the vacuum state as the input to Alice's OPA, the output state after Bob's OPA is 
\begin{equation}
\label{equation:qkd_quantum_state}
\left|\psi\right>_{2} = 
\left|0\right> + 
ig_{\omega} (1 + e^{i\phi_{\omega}}) \left|1_{\omega}, 1_{-\omega}\right>,
\end{equation}
where $\phi_{\omega}=\phi_{A} + \phi_{B}$ is the total phase that Alice and Bob apply to the signal-idler pair (relative to the pump field) during steps 1 and 2 of the protocol. The phase $\phi_{A(B)}$ indicates the basis of encoding (measurement) that Alice (Bob) employ for each bit of the channel. The average number of photons that Bob will measure at a specific channel is
\begin{equation}
\label{equation:qkd_photon_number}
N_{\omega} = |\left<1_\omega|\psi\right>_{2}|^{2} = 
|g_{\omega}|^{2}(2 + 2\cos(\phi_{\omega})),
\end{equation}
where for simplicity we assumed that Bob's parametric gain is identical to Alice (generalization to nonidentical gain is simple, see \cite{Santandrea_2023}). Notice that when Alice and Bob use different bases, $\phi_{\omega} = \phi_{A} + \phi_{B} = \frac{\pi}{2} / \frac{3\pi}{2} $ and the average number of photons is simply $2|g_{\omega}|^{2}$, independent of the phase. However, when the bases match, $\phi_{\omega} = \phi_{A} + \phi_{B} = \pi / 0$ and the average number of photon equals $0$ or $4|g_{\omega}|^{2}$ respectively, so Bob's measurement can decode Alice's information. 

Our analysis so far assumed the parametric gain to be low and neglected the generation of multiple photon pairs per bit. Generally speaking, multiple pairs are not desirable, as they open the door to splitting attacks, but note that multiple pairs do not hamper the communication protocol or introduce errors. Indeed, the phase of each frequency (or photon) is random and varies with time, but the phase-sum of  the signal-idler pair is deterministically fixed by the pump (and propagation). Since our protocol relies only on the phase-sum, no phase ambiguity will arise, even in the case of multiple photons per bit.

When Eve tries to attack the communication the situation changes noticeably. In the supplementary material we analyze some of the possible realistic attacks that Eve may try. However, in order to illustrate Eve's detectability, let us consider the simplest possible example - the steal attack. If Eve "steals" some of the light using a beam splitter with transmission $T$, then the number of photons after Bob's OPA becomes (see derivation in supplement 1)
\begin{equation}
\label{equation:steal_attack_number_of_photon}
N_{\omega}(T) = 
|g_{\omega}| ^ {2} \left(1 + T + 2T\cos(\phi_{\omega})\right)
\end{equation}
which diminishes the interference contrast and introduces errors for Bob. Thus, a good discriminator for eavesdroppers is the contrast of the interference after Bob's crystal,
\begin{equation}
\label{equation:steal_attack_contrast}
V(T) \equiv \frac{I_{max} - I_{min}}{I_{max} + I_{min}} = \frac{2T}{1+T},
\end{equation}
where again the parametric gains of Alice and Bob are assumed equal (for non equal gains the visibility is $V(T) = \frac{2g_{A}g_{B}T}{g^{2}_{B}+Tg^{2}_{A}}$ \cite{Santandrea_2023} ). The contrast is a witness for eavesdropping since Eve must steal some of the photons, i.e. reduce the transmission ($T$) which will lower the contrast. Notice that Eve's ability to extract information relies on a similar interference contrast (in her own measurements), $V_{Eve}=V(R=1\!-\!T)=\frac{2(1-T)}{2-T}$. Thus, to obtain a sufficient contrast in her measurements, Eve must induce a sufficiently high loss in her beam splitter, which Alice and Bob can later identify. Although the steal attack above is simplistic, and may even be detected by measuring the average intensity, the contrast proves to be a very effective discriminator for Eve's presence, even with much more realistic and sophisticated attacks (some of which are analyzed the supplementary material).

It is important to understand that our protocol encapsulates a set of single QKD communication channels that are otherwise independent, which indicates that standard QKD constraints apply at the single channel level and not at the entire protocol level. As an example, our spectrally multiplexed QKD allows for a much higher total photon flux (and much higher total data rate) than the equivalent single channel QKD. Specifically, for QKD to be secure, each transmitted bit should contain less than one photon per detection time, but since the spectral channels carry the QKD exchange independently, this low-flux requirement holds \textbf{per channel}. Thus, the total photon generation rate (parametric gain) should be adjusted to produce less than one photon on average is sent \textbf{per channel bandwidth} per detection time. The total flux of photons across all the channels can be much more than one photon.

Clearly our protocol requires Alice and Bob to share a classical phase reference and to be able to modulate the spectral phase relative to it. While figure \ref{fig:QKD_scheme} shows one possible implementation to meet these requirements (by sending the pump along with the SPDC as a common phase reference and using a Fourier domain pulse shaper as a spectral phase modulator), other implementations are possible that may prove technically advantageous in specific configurations. We will briefly consider some of those possibilities in the discussion section later on.

\section{Multiplexed Quantum Teleportation}
\label{section:quantum_teleportation}

\begin{figure}[ht!]
\centering\includegraphics[width=\textwidth]{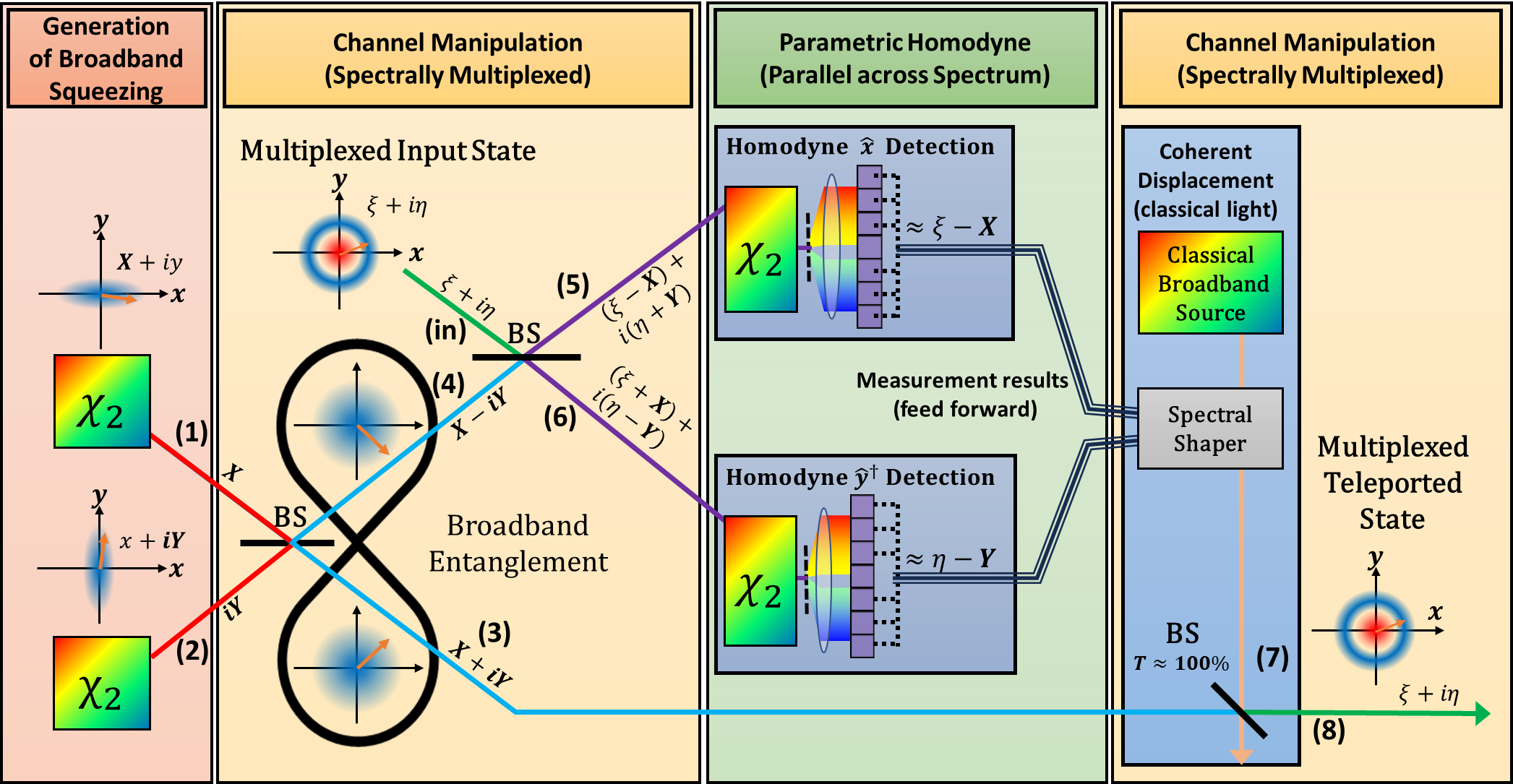}
\caption{\textbf{The multiplexed quantum teleportation protocol}. Two OPAs are generating two broadband orthogonal squeezed states, one in the $\hat{x}$, the other in $\hat{y}$ directions ($X,iY$ marked (1) and(2)). These states are interfered using a beam splitter to produce two quadrature-entangled states ($X\pm iY$ marked (3 and (4)). One of the entangled states (3) is sent on to the teleporting site, while the other (4) is mixed on a beam splitter (BS) with the input signal to be teleported ($\xi+i\eta$, marked (in)), which produces two interference outputs (marked (5) and (6)) that are them measuredby two orthogonal broadband parametric homodyne detectors - one to measure the $\hat{x}_\omega$ quadrature and the other $\hat{y}^\dagger_\omega$ across the spectrum, which provides the instantaneous difference between the entangled state quadratures $\xi-X, \eta-Y$ without measuring $\xi,X,\eta,Y$ themselves. Finally, these measurement results are communicated classically (over a classical channel, marked thick black) to the teleporting site, where it is used to generate a coherent classical local oscillator $\alpha_{\text{shift}}(\omega)\sim\xi-X +i\eta-Y$ (using a spectrally shaped broadband classical source / ultrashort pulse). This LO $\alpha_{\text{shift}}(\omega)$ is then mixed on an unbalanced beam-splitter with the received state (3) in order to shift it by the exact amount needed to reproduce the input state $\xi+i\eta$  at the output (8) on every frequency channel.}
\label{fig:quantum_teleportation_scheme}
\end{figure}

Multi-channel quantum teleportation, which we propose and analyze here, is another example for harnessing the optical bandwidth to multiplex an important protocol of quantum information. Again we note that this protocol should not be judged as immediately applicable for technology, but rather as an illustration of the range of possibilities that our multiplexing scheme offers for quantum information. This new protocol is a broadband, multiplexed version of the Braunstein's \& Kimble's protocol (suggested in \cite{BraunsteinKimble} and demonstrated in \cite{TeleportationDemonstration}). For this scheme we utilize two sources of broadband two-mode squeezed vacuum (OPAs) to simultaneously teleport a set of arbitrary quantum states across the spectrum of a general broadband field (see figure \ref{fig:quantum_teleportation_scheme}). 

Before delving into the details of the teleportation protocol, let us briefly review the continuous-variable description of quantum optics with broadband fields using two-mode quadratures (rather than the discrete basis of photon number). We adopt the formalism of two-mode quadratures $\hat{x}_{\omega} \equiv \frac{1}{2} \left(\hat{a}_{\omega} + \hat{a}_{-\omega}^\dagger\right)$ and $\hat{y}_{\omega}\equiv \frac{i}{2} \left(\hat{a}_{\omega}^\dagger-\hat{a}_{-\omega}\right)$ \cite{TwoModeQuadrature, TwoModeQuadrature2}, where the $\omega,-\omega$  indices denote the signal and idler frequencies relative the the center carrier frequency of $\omega_p/2$ (later to become independent spectral channels of teleportation). The two mode quadratures are a generalization of the well known single-mode quadratures that enables to treat squeezing on equal footings, be it degenerate and non-degenerate (two-mode) squeezing under a unified framework (maintaining the canonical commutation $\left[\hat{x}_\omega, \hat{y}_\omega\right]\!=\!\frac{i}{2}$) . 

With two-mode quadratures we can decompose the field operator $\hat{a}_{\omega}$ at any frequency $\omega$ as $\hat{a}_{\omega} \!=\! \hat{x}_{\omega} + i\hat{y}_{\omega}^\dagger$ . Adopting the Heisenberg picture of quantum dynamics, the field operators evolve in time under the Hamiltonian of the system. Thus, the field operator of a squeezed state generated by an OPA is $\hat{a}_{\omega,OPA} = e^g\hat{x}_{\omega} + ie^{-g}\hat{y}_{\omega}^{\dagger}= X(\omega)\hat{x}_{\omega} + iy(\omega)\hat{y}_{\omega}^{\dagger}$, where $X\!=\!e^g$ represents the stretched two-mode quadrature and $y\!=\!e^{-g}$ the squeezed ($g(\omega)$ is the parametric gain).  Similarly, we can represent the field operator of the input state of channel $\omega$ (to be teleport) as $\hat{a}_{\omega,in} = \xi(\omega)\hat{x}_{\omega} + i\eta(\omega)\hat{y}_{\omega}^{\dagger}$, where the teleportation task is to transfer both quadratures $\xi(\omega),\eta(\omega)$ from the input to the output without measuring any one of them. For convenience, we will drop the frequency index from now on (assuming we look at a specific frequency if component of the broad spectrum). 
 
As opposed to the QKD protocol above, which operates in the regime of low parametric gain, where the squeezing is negligible and the generated light can be considered as a random stream of  single bi-photons, the teleportation protocol here operates best at the high parametric gain regime of multiple photons and high-squeezing. Let us therefore assume (for simplicity of the presentation) that the two sources are "infinitely squeezed", such that we can completely ignore the squeezed quadratures. Later we will alleviate this assumption and consider the implications of finite squeezing to the teleportation precision. 

Let us now describe in detail each step of this protocol, which is also visually presented in figure \ref{fig:quantum_teleportation_scheme}:
\begin{enumerate}
    \item Two broadband squeezed sources generate two orthogonal squeezed states, $\hat{a}_{1} = \sqrt{2}\left(X\hat{x} + iy\hat{y}^\dagger\right)$ and $\hat{a}_{2} = \sqrt{2}\left(x\hat{x} + iY\hat{y}^\dagger\right)$, where $X,Y$ ($y,x$) are the stretched (squeezed) quadratures of the sources. We will assume that the squeezing is sufficiently high to ensure that the squeezed quadratures are small compared to the input field, i.e. $x \ll \xi$ and $y \ll \eta$.

    \item Using a beam splitter, the squeezed states are interfered to generate two quadratures-entangled states, $\hat{a}_{3} = \frac{1}{\sqrt{2}}\left(\hat{a}_{1} + \hat{a}_{2}\right) \approx X\hat{x} + iY\hat{y}^\dagger$ and $\hat{a}_{4}  = \frac{1}{\sqrt{2}}\left(\hat{a}_{1} - \hat{a}_{2}\right) \approx X\hat{x} - iY\hat{y}^\dagger$, where $x,y$ are neglected for now. 
    %The entangled state $\hat{a}_{\omega,3}$ will be measured against the input state, whereas $\hat{a}_{\omega,4}$ is transmitted to the desired teleportation location.

    \item \label{step:input_signal_step}The broadband input state that we wish to teleport, represented by the field operator $\hat{a}_{in} = \xi\hat{x} + i\eta\hat{y}^\dagger$, is mixed with one of the entangled beams $\hat{a}_{4}$ using a second beam splitter to obtain the encoded states, $\hat{a}_{5} \approx \frac{1}{\sqrt{2}}\left(\xi-X\right)\hat{x} + \frac{i}{\sqrt{2}}\left(\eta+Y\right)\hat{y}^\dagger$ and $\hat{a}_{6} \approx \frac{1}{\sqrt{2}}\left(\xi+X\right)\hat{x} + \frac{i}{\sqrt{2}}\left(\eta-Y\right)\hat{y}^\dagger$. 

    \item The quadratures of the two encoded states are measured (intensity and phase) with parametric homodyne measurement\cite{OpticalParametricHomodyne, Li:19, PhysRevA.101.053801, 10.1063/5.0137641} simultaneously across the two-mode spectrum, such that the quadrature $\hat{x}$ is measured for $\hat{a}_{5}$ and $\hat{y}^\dagger$ for $\hat{a}_{6}$. As a result, we obtain information on \textit{the difference} of the signals quadrature, $\frac{1}{\sqrt{2}}\left(\xi-X\right)$ and $\frac{1}{\sqrt{2}}\left(\eta-Y\right)$ without any knowledge about the quadratures themselves. 

    \item The measurement results of the quadratures are transmitted through a classical channel to the desired teleportation location, where a strong coherent state (effectively classical) is generated from the received measurements (using spectral shaper) according to $\hat{a}_{7} \approx \alpha\left(\xi-X\right)\hat{x} + i\alpha\left(\eta-Y\right)\hat{y}^\dagger$. To recreate the original input state at the teleportation output, we use this coherent state to shift the quadratures of the remaining part of the entangled state, $\hat{a}_{3}$ using a beam splitter with high-transmission ($t \approx 1, r \ll 1$). To this end, we set $\alpha = \frac{t}{r}$, which yields $\hat{a}_{8} = t\hat{a}_{3} + r\hat{a}_{7} \approx \xi\hat{x} + i\eta\hat{y}^\dagger = \hat{a}_{in}$. \label{step:num_teleport_steps} 
\end{enumerate}

It is important to note that the level of squeezing of the OPAs is a key factor for the fidelity of the protocol, indicating that the teleportation error is a direct result of the finite squeezing used. If we calculate the output of the protocol $\hat{a}_{8}$  without assuming high squeezing, i.e. including also the squeezed quadratures of the OPAs ($x, y$), the output operator becomes 
\begin{equation}
    \hat{a}_{8} = t\left((2x + \xi)\hat{x} + i(2y + \eta)\hat{y}^\dagger\right) \approx (2x + \xi)\hat{x} + i(2y + \eta)\hat{y}^\dagger,
\end{equation}
indicating that the residue of the squeezed quadratures acts as a source of noise, added to the teleportation output. Thus, in order to reduce these errors, it is important to maximize the squeezing of the original signals and to minimize the loss of the transmitted quantum state, $\hat{a}_{3}$ (as loss reduces squeezing and introduces vacuum noise). For example, one can enhance the squeezing level by replacing the single-pass OPAs in our protocol with multi-pass OPOs (optical parametric oscillator) that offer higher squeezing (up to 15 dB demonstrated \cite{Schnabel}).

Just like the QKD protocol above, the teleportation requires a shared classical resource - a phase reference. While one possible implementation is to send the pump along with the squeezed spectrum, many other options can be conceived according to the technical limitations of the classical communication. We highlight other possibilities in the discussion below.

\section{Experimental Demonstration}
\label{section:multiplexed_qkd_experiment}
\begin{figure}[ht!]
\centering\includegraphics[width=10.5cm]{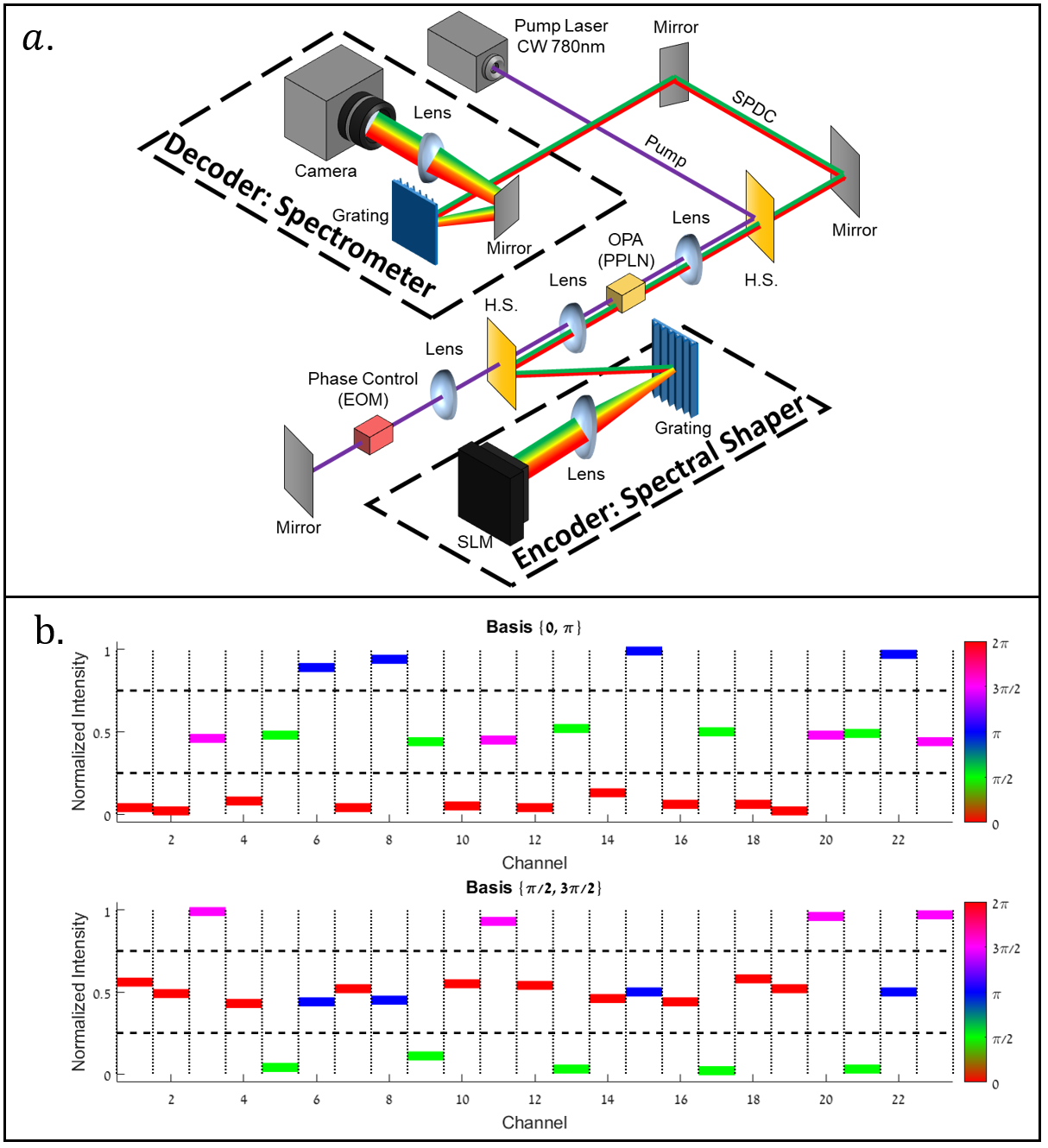}    
\caption{\scriptsize\textbf{Proof-of-principle demonstration of multiplexed QKD}: (a) Experimental configuration The pump (purple) was a CW, single-frequency laser at $780nm$ of $\sim 1-4W$ power. The two OPAs of the SU1,1 interferometer were realized in a single periodically polled LiNbO3 crystal (PPLN), by passing through the crystal twice: forward for generation of the broadband bi-photons (red and green); and backwards, for parametric homodyne detection. Alice encodes the bi-photons spectrum using a spectral shaper and a phase modulator (EOM) for the pump, which is separated off from the bi-photons by a harmonic separator (H.S). Alice encodes information with a spectral shaper that consists of a grating, a lens and a spatial light modulator (SLM - 1x12,288 linear array of liquid crystal pixels). The EOM stabilizes the phase of the pump to that of the bi-photons, compensating for phase noise and drifts with an active feedback loop (not shown). The EOM also allows Bob to select the measurement basis (same for all channels across the spectrum). Finally, the three beams are reflected back into the OPA crystal, which now acts as Bob's detector. The output bi-photons beam is separated from the pump again with another H.S mirror and the resulting bi-photons spectrum is measured on a home-built spectrometer that consists of a grating, a lens and a Linear CCD camera. (b) Experimental results: the normalized  spectral intensities of all channels, which represent the detection probability of a photon per channel, are shown for measurement in basis "1" (top) and "2" (bottom). The colors represents the encoded phase by Alice for each channel, and the height is the normalized intensity (detection probability) measured by Bob. Evidently, Bob can easily decode only the channels that were measured at the correct basis.}
\label{fig:experimental_setup}
\end{figure}

To demonstrate multiplexed quantum information processing across the optical spectrum we implemented the multi-channel QKD scheme of figure \ref{fig:QKD_scheme} in a proof-of-principle experiment, as outlined in figure \ref{fig:experimental_setup}. Our configuration realized the simultaneous generation, control and measurement of multiplexed QKD frequency channels. The experimental setup is illustrated in figure \ref{fig:experimental_setup}a.

During the experiment the pump passes through the OPA and generates a broadband spectrum of signal and idler pairs across $~150nm$ bandwidth around $1560nm$. The pairs are separated from the pump with a Harmonic Separator (H.S) and sent to a spectral shaper to encode Alice's information by modulating the phase of each frequency pair (channel) to $0, \pi$ or $\frac{\pi}{2}, \frac{3\pi}{2}$ randomly (step 1). The spectral shaper also serves to compensate for the chromatic dispersion that the signal and idler accumulate during the propagation through the optical configuration between the OPAs. Simultaneously, the pump passes through a phase modulator (EOM) used to stabilize the phase of the pump relative to that of the pairs, as well as to choose the measurement basis for Bob  (same basis for all channels in this case). Finally, the pump and the pairs are reflected back to recombine and pass through the OPA once more in the opposite direction, completing the SU(1,1) interference. To implement Bob's parallel detection of all channels we measure the output spectral intensity using a home-built spectrometer composed of a grating and a line CCD-camera (step 2). We tuned Bob's parametric gain to roughly equal that of Alice by polarization manipulations on the pump beam (not shown). Practically, this could also be achieved by classical amplification, since the gain (pump) is a classical resource.

To maximize the data capacity while preserving the security of the protocol, the spectral width of the channels at the spectral shaper was chosen to be the smallest possible without leaking to the neighboring channels (see experimental verification in supplement 1). This allowed us to encode and decode in this preliminary configuration up to 23 channels in parallel with a bare interference contrast of $75\%$. This 23-fold enhancement of the data capacity (compared to a single channel) in this proof-of-principle experiment is far from any fundamental limit for two reasons. First, the number of channels can easily be pushed up to $>\!200$ channels by improving the spectral resolution of both the encoding shaper and the spectrometer with standard technology of optical wavelength division multiplexing (see supplement 1). Second, the low communication speed of the single channel ( $\sim\!100$Hz due to the slow speed of the liquid crystal SLM) can be drastically uplifted with real-time phase modulators (electro-optic or acousto-optic), which can reach up to GHz rates.

For QKD to operate securely, it is preferable to maintain a low parametric gain with $N_q\ll1$ photons per quantum state. The bandwidth of each channel in our experiment was $\sim100GHz$, which indicates a coherence time of $\tau_c\sim\frac{1}{100GHz} = 10ps$ per channel, which was much longer than the phase delay induced by the SLM (of order 1 optical cycle, maximum), certifying ideal visibility \cite{TemporalCoherence, QuantumOpticalCoherence}. The pump power in our experiment ($4W$) generated a flux of up to $10^7$ photons/sec per channel (as measured on our CCD), which indicates $N_q\leq 10^{-4}$ per quantum state - safely within the low gain regime. The integration time per bit in the QKD protocol can then be chosen at $10^{4}\tau_c = 100ns$, such that the average detection probability of a photon will be near $N_{detect}\sim 1$ per bit. To demonstrate that the interference outcome is easily identified in our experiment for the different bases, we purposely used a much longer integration time on our CCD (about $1ms$) with $\sim10^4$  photo-detection events per integration time, where the measured intensity directly represents the probability of detection for every channel.

Figure \ref{fig:experimental_setup}b presents the experimental results. To demonstrate our ability to decode the information in both bases, we encoded the 23 channels (at random bases) using the spectral shaper and then, measured them all simultaneously, where in order to select the measurement basis we set the pump-phase to $0$ or $\frac{\pi}{2}$. As can be seen, for each measurement the correct basis was detected with good visibility across the entire spectrum, allowing to decode the channels, whereas the channels in the incorrect basis showed no visibility at all. This confirms the ability of Alice and Bob to communicate freely, while preventing attacks on the communication by intercept-resend.

As discussed above, our scheme is a CV, multiplexed version of the well-known BB84 protocol, indicating that all the security proofs of BB84 are directly applicable to our protocol as well. We chose to demonstrate the immunity of our scheme to the steal-attack, i.e. an attempt of Eve to split off some of the light between Alice and Bob, which is a most common attack on QKD. If Eve uses a beam splitter to "steal" part of the quantum state, she inevitably will reduce the interference contrast for Bob. We therefore simulated Eve's operation by introducing loss at the spectral shaper. Although the initial loss in our simple configuration was relatively high ($~44\%$, probably due to imperfect components and alignment), we could still clearly detect even a small additional loss of $5\%$ since it visibly reduces the contrast (compared to the measurement error). For more details about the loss detection see supplement 1.

\section{Discussion}
To perform quantum information processing with our tools, some classical and technical resources are required, such as the classical phase reference and the spectral phase modulation device. In the description of our protocols above and in our experimental demonstration we chose simple implementations of these resources: sending the pump as a phase reference and using a Fourier domain shaper as the broadband phase modulator. However, many other implementations are possible that may be advantageous for specific realizations under given technical constraints. 

For example, the update rate of the liquid-crystal based SLM in our spectral shaper is rather slow ($\sim100$Hz), which would limit the data rate per channel to be impractically low for a realistic QKD application. In addition, the transmission of the fiber connecting the communicating parties may be too lossy in a real scenario to transmit the pump laser from Alice to Bob; and the chromatic dispersion of that fiber may require handling / compensation that is beyond the capability and dynamic range of the spectral shaper. 

While these technical considerations are very important for a practical realization, they are classical constraints that can be overcome in various classical means, depending on the technical details of the implementation. For example, instead of the pump one can utilize the degenerate frequency channel at $\omega_p/2$ as phase reference (the center of the SPDC spectrum) to send a laser as a classical local oscillator that is less attenuated in the connecting fiber (and can be classically amplified, if needed). Other methods for distributing a time-base across the fiber can be adequate as well \cite{PhaseSynchronization1,PhaseSynchronization2,PhaseSynchronization3}. Fast spectral modulation can be implemented using arrays of electro-optic modulators \cite{LiNoSpatialModulator} or fast acoustic modulators \cite{AOMSpatialModulator}, or even internally in the fiber with a space-division-multiplexing networks \cite{WDM-SDM}. Chromatic dispersion can be compensated either by spectral modulators (as we did in our experiment above) or by incorporating a dispersion compensating fiber. 

\section{Conclusion}
We presented a set of tools to efficiently multiplex quantum information processing across a broad bandwidth of squeezed light with a large number of frequency channels. To demonstrate the capabilities of these tools, we developed multiplexed versions for two of the most important quantum communication protocols: QKD, which was also demonstrated in a proof-of-principle experiment on 23 parallel channels and quantum teleportation. Since broadband sources of squeezed light and entangled photons can easily exceed $10$THz (up to an octave in frequency \cite{OctaveSqueezedLight}), our new method can potentially speed-up quantum processing by orders of magnitude compared to the standard, single channel implementations.

In a broader view, our set of tools is applicable to quantum processing in general, and can be used as building blocks for frequency multiplexed versions of many other quantum protocols, such as quantum coin flipping \cite{BENNETT20147}, entanglement-based QKD \cite{Eckert91} and entanglement based sensing \cite{EntanglementSensing}, which we plan to pursue in the future. Furthermore, our tools can be extended to include two-qubit operations (that have already been demonstrated across the quantum frequency comb of a broadband OPO \cite{MultimodePump, PhaseModulationClusterState}), which can pave the way towards high-bandwidth quantum computation with the potential to be fast, scalable (in the number of qubits) and compatible with broadband quantum communication networks \cite{MultimodePump, OneWayQuantumComputing, Pfister_2020}.

 \begin{backmatter}

\bmsection{Acknowledgements}
This research was funded in part by SPARQL consortium, under the QuantERA program of the EU.

\bmsection{Supplemental document}
See Supplementary Material for supporting content. 

\end{backmatter}

\bibliography{bibliography/bibliography}

\end{document}

\title{Supplementary Material}

\maketitle

\section{Security Analysis of Frequency-Multiplexed QKD}
\label{section:security_analysis}

In the main text (section 2), we proposed a frequency-multiplexed QKD scheme that employs broadband squeezing and broadband quantum detection, and described its use to share information securely between Alice and Bob using the phases of multiple frequency channels. The security of these channels relies on the ability to efficiently detect eavesdroppers, which requires to identify a difference between the quantum state with and without an eavesdropper (and to measure this difference). For an undisturbed system, the output state is given by equation 1 (section 2), 
\begin{equation}
\left|\psi\right>_{2} = 
\left|0\right> + 
ig_{\omega} (1 + e^{i(\phi_{A} + \phi_{B})}) \left|1_{\omega}, 1_{-\omega}\right>,
\end{equation}
which yields the detection-probability of photons (phase-dependent) in mode $\omega$, as given by equation 2 of the main text (section 2),
\begin{equation}
N_{\omega} = 
|g_{\omega}|^{2}(2 + 2\cos(\phi_{A} + \phi_{B})).
\end{equation}
In what follows, we analyze how this output measurement will change under two major attacks - the steal attack and the intercept-resend attack, and how these attacks will be reflected in the statistics of the results. 

\subsection{Intercept-Resend Attack}
\label{section:Intercept-Resend Attack}
Intercept-resend attack was introduced originally with the BB84 protocol. During an intercept-resend attack, Eve tries to imitate Alice by reading the quantum state between Alice and Bob and generating a new quantum state according to her measurement, that she sends to Bob.

Following the calculation in section 2 (main text), the average number of photons that Eve will measure after her OPA is
\begin{equation}
N_{\omega} = |g_{\omega}|^{2}(2 + 2\cos(\phi_{A} + \phi_{E})),
\end{equation}
where $\phi_{E}$ is the modulated phase by Eve.

We can gain intuition to the limitations of the intercept-resend attack by considering the simplifying assumption that the number of measured photons during each integration time is exactly 1 (in every shot) for constructive interference (and zero for destructive). In this case, Eve cannot gain any information about Alice's basis, since her measurements will always yield either one photon or none, independent of Alice's encoding. Now, exactly as in BB84, if Eve measured in the correct basis, she will know the state correctly and will be able to impersonate Alice perfectly. However, if Eve measured in the incorrect basis, her reading is random and she cannot recover the encoded phase, thereby introducing an error to the channel with probability $0.5$. The total error probability per bit is therefore $P_{Err} = 0.25$, as in BB84, which can be detected easily.

\subsection{Steal Attack}
\label{section:steal_attack}
One may claim that the intercept-resend attack fails because it attempts to obtain the complete information between Alice and Bob via a full interception of the communicated light. Thus, one may suggest that a more subtle attack, where Eve obtains only a fraction of the information, but with a lower signature that Alice and Bob may not be able to detect.

Under a steal attack, Eve tries to "steal" a fraction of the transmitted light by a beam splitter in the channel with a reflection $R$ representing the stolen amplitude. Eve can then try to use her stolen part to obtain partial  information on the generated key. Let us first assume a simple case, where Eve only steals some of the quantum state, but does not replenish it with new light. Indeed, such a simple steal allows Alice and Bob to detect Eve's influence by classical means, simply by noticing that the average intensity through the channel has dropped, which  seems to defeat the original purpose of an attack on the protocol. However, the analysis of this "simpler" attack is a convenient stepping stone for the analysis of the more sophisticated version - the steal-intercept attack, given in the next section.

During steal attack, Eve modifies the transmitted state to Bob by mixing it with another vacuum mode $\left|0\right>_2$ through the beam-splitter, resulting in:
\begin{equation}
\left|\psi\right>_{BS} = 
\left|0\right>_{1} \left|0\right>_{2} + 
ig_{\omega}e^{i\phi_{A}}
\begin{bmatrix}
	 t^{2} \left|1_{\omega}, 1_{-\omega}\right>_{1} \left|0\right>_{2} \\
	- r^{2} \left|0\right>_{1} \left|1_{\omega}, 1_{-\omega}\right>_{2} \\
	+irt \left|1_{\omega}, 0_{-\omega}\right>_{1} \left|0_{\omega}, 1_{-\omega}\right>_{2} \\
	+irt \left|0_{\omega}, 1_{-\omega}\right>_{1} \left|1_{\omega}, 0_{-\omega}\right>_{2}
\end{bmatrix},
\end{equation}
where $t,r$ are the transmission and reflection amplitudes of Eve's beam-splitter ($t^2+r^2=1$).

Then, the quantum state after Bob's OPA becomes:
\begin{equation}
\left|\psi\right>_{BS} = 
\left|0\right>_{1} \left|0\right>_{2} + 
ig_{\omega}e^{i(\phi_{A} + \phi_{B})}
\begin{bmatrix}
	(e^{-i(\phi_{A} + \phi_{B})} + t^{2}) \left|1_{\omega}, 1_{-\omega}\right>_{1} \left|0\right>_{2} \\
	- r^{2} \left|0\right>_{1} \left|1_{\omega}, 1_{-\omega}\right>_{2} \\
	+irt \left|1_{\omega}, 0_{-\omega}\right>_{1} \left|0_{\omega}, 1_{-\omega}\right>_{2} \\
	+irt \left|0_{\omega}, 1_{-\omega}\right>_{1} \left|1_{\omega}, 0_{-\omega}\right>_{2}
\end{bmatrix},
\end{equation} 
where the modulation phase, $\phi_{B}$ sets Bob's measurement basis. The average number of photons that Bob will measure in mode $\omega$ is:
\begin{equation}
N_{\omega} = |\left<1_\omega|\psi\right>_{BS}|^{2} = 
|g_{\omega}| ^ {2} \left(1 + T + 2T\cos(\phi_{\omega})\right)
\end{equation}
where $T = t^2, R=r^2$ are the transmission / reflection probabilities.

\begin{equation}
N_{\omega} = |\left<1_\omega|\psi\right>_{BS}|^{2} = 
|g_{\omega}| ^ {2} \left(1+ T + 2T\cos(\phi_{\omega})\right)
\end{equation}
The primary method to detect Eve is through the errors she will induce in the measurements, which appear in two major forms: First, Eve's beam-splitter may steal one photon of an entangled pair, but not the other, which leads to the observation of single photons without a matching twin (i.e. the states $\left|1_{\omega}, 0_{-\omega}\right>$ and $\left|0_{\omega}, 1_{-\omega}\right>$). Since these possibilities cannot exist in the ideal case, they are good indicators for the amount of loss in the communication channel, which we attribute to Eve. The second type of errors is the "information" errors, where the outcome of legitimate measurements is altered. Specifically, photons can be detected even when the interference is destructive, which reduces the interference contrast. Thus, evaluating the contrast of a large set of measurements is a good discriminator, which is given by
\begin{equation}
\label{equation:scheme_contrast}
V \equiv \frac{I_{max} - I_{min}}{I_{max} + I_{min}} = \frac{2T}{1+T}.
\end{equation}
The contrast is a witness for eavesdropping since Eve must steal some of the photons, i.e. reduce the transmission ($T$) which will lower the contrast. Additionally, we can notice that a background signal (such as noises and signals generated by the attacker) will reduce the contrast even further, and so reduce the signal's credibility. Notice that Bob and Alice can easily calculate both $I_{max}$ and $I_{min}$ during the comparison step of our QKD protocol (section 2, step 4). Specifically, when Alice reveals the transmitted states of the compared data, this allows Bob to calculate the average number of photons that he received for constructive (destructive) interference and obtain $I_{max}$ ($I_{min}$). Note that for Eve to obtain useful information, she must steal a substantial fraction of the light, since the transmission of her beam splitter (from Alice to Bob) acts as the loss value for Eve. For example, stealing $5\%$ is equivalent to $95\%$ loss for Eve, which results in only $\sim10\%$ contrast for Eve and provides little information. 

\subsection{Steal-Resend Attack}
\label{section:Steal-Resend Attack}
During a steal-resend attack Eve steals part of the quantum state, similar to steal attack and attempts to cover her tracks by generating a new quantum state (similar to the generated state in the intercept-resend attack) and sending it to Bob together with the remainder of the original state. While this prevents Eve from being detected by classical means, it still hampers the detection contrast, which still serves as a good discriminator for Alice and Bob to uncover the eavesdropper.

Similar to steal attack, we can model the "steal" operation using a beam splitter with loss $R$ representing the stolen part by Eve. Here the quantum state after the beam splitter is given by: 
\begin{equation}
\left|\psi\right>_{B.S.} = 
\left|0\right>_{Bob} \left|0\right>_{Eve} + 
ig_{\omega}e^{i\phi_{A}}
\begin{bmatrix}
	 t^{2} \left|1_{\omega}, 1_{-\omega}\right>_{Bob} \left|0\right>_{Eve} \\
	- r^{2} \left|0\right>_{Bob} \left|1_{\omega}, 1_{-\omega}\right>_{Eve} \\
	+irt \left|1_{\omega}, 0_{-\omega}\right>_{Bob} \left|0_{\omega}, 1_{-\omega}\right>_{Eve} \\
	+irt \left|0_{\omega}, 1_{-\omega}\right>_{Bob} \left|1_{\omega}, 0_{-\omega}\right>_{Eve}
\end{bmatrix}.
\end{equation}
Eve takes the reflected beam of the beam splitter to be measured ("Eve") and the transmitted beam can be later modified and sent further to Bob ("Bob"). From Eve's perspective, the stolen state is very similar to the quantum state during the simple steal attack above, but the reflection $r$ and the transmission $t$ switch roles, since Eve receives the stolen part and "loses" the transmitted light. Because Eve does not know Alice's basis of measurement, she measures her stolen quantum state in a random basis, adding a random phase $\phi_{E} = 0 \text{ or } \frac{\pi}{2}$ to the light, similar to Bob. After Eve's parametric amplifier, her state becomes
\begin{equation}
\left|\psi\right>_{B.S.} = 
\left|0\right>_{Bob} \left|0\right>_{Eve} + 
ig_{\omega}e^{i\phi_{AE}}
\begin{bmatrix}
	t^{2} \left|1_{\omega}, 1_{-\omega}\right>_{Bob} \left|0\right>_{Eve} \\
	+(e^{-i\phi_{AE}} - r^{2}) \left|0\right>_{Bob} \left|1_{\omega}, 1_{-\omega}\right>_{Eve} \\
	+irt \left|1_{\omega}, 0_{-\omega}\right>_{Bob} \left|0_{\omega}, 1_{-\omega}\right>_{Eve} \\
	+irt \left|0_{\omega}, 1_{-\omega}\right>_{Bob} \left|1_{\omega}, 0_{-\omega}\right>_{Eve}
\end{bmatrix},
\end{equation}
where $\phi_{AE} = \phi_{A} + \phi_{E}$. 

After the parametric amplifier, Eve measures her state to gain information about Alice's state. This measurement has several possible outcomes: A photon pair (signal and idler); single photon in only one of the signal or idler; or no photons at all. The probabilities for these outcomes are summarised in  table \ref{table:photon_measurement_ratios}.

\begin{center}
	\begin{tabular}{c  |c llc}
		$P(N_\omega, N_{-\omega})$& $\phi_{AE} = 0$& $\phi_{AE} = \frac{\pi}{2}$&$\phi_{AE} = \pi$& $\phi_{AE} = -\frac{\pi}{2}$\\
		\hline
		$P(1,1)$ & $\frac{1}{\mathcal{Z}_{0}}(R+1)^2$& $\frac{1}{\mathcal{Z}_{\frac{\pi}{2}}}|R-i|^2$&$\frac{1}{\mathcal{Z}_{\pi}}(R-1)^2$& $\frac{1}{\mathcal{Z}_{-\frac{\pi}{2}}}|R+i|^2$\\
		$P(1,0) + P(0,1)$ & $\frac{1}{\mathcal{Z}_{0}}2RT$  & $\frac{1}{\mathcal{Z}_{\frac{\pi}{2}}}2RT$&$\frac{1}{\mathcal{Z}_{\pi}}2RT$& $\frac{1}{\mathcal{Z}_{-\frac{\pi}{2}}}2RT$\\
		$P(0,0)$ & $\frac{1}{\mathcal{Z}_{0}}\left(T - \frac{1}{ g_{\omega}}\right)^2$& $\frac{1}{\mathcal{Z}_{\frac{\pi}{2}}}\left( T -\frac{1}{ g_{\omega}}\right)^2$&$\frac{1}{\mathcal{Z}_{\pi}}\left(T - \frac{1}{ g_{\omega}}\right)^2$& $\frac{1}{\mathcal{Z}_{-\frac{\pi}{2}}}\left( T -\frac{1}{ g_{\omega}}\right)^2$\\
	\end{tabular}
	\captionof{table}{The possible outcomes of Eve's measurements and their probabilities for all the possible phase selections $\phi_{AE}$ ($0, \pi$ and $\pm\frac{\pi}{2}$): $P(1,1)$ - photon pair in both the signal and the idler, $P(1,0) + P(0,1)$ - photon in only one of them, and $P(0,0)$ - no photons at all. $\mathcal{Z}_{*}$ are the relevant normalization factors for each $\phi_{AE}$. Notice that the distribution of measurements in the wrong basis, where $\phi_{AE} = \pm\frac{\pi}{2}$ are identical, meaning the Eve cannot deduce anything on Alice's basis, as expected.}
	\label{table:photon_measurement_ratios}
\end{center}

When Eve measures at the wrong basis, which happens at 50\% of the cases on average, $\phi_{AE} = \pm\frac{\pi}{2}$ with identical measurement distributions. Thus, Eve cannot deduce Alice's phase from her measurement. However, in the correct basis, where $\phi_{AE} = 0 \text{ or } \pi$ the measurement probabilities are different and Eve can guess Alice's phase with better than random success. Specifically, since $\left. P(1,1) \right|_{\phi_{AE} = 0} > \left. P(1,1) \right|_{\phi_{AE} = \pi}$, Eve can guess $\phi_{AE} = 0$ when she measures a photon pair and $\phi_{AE} = \pi$ otherwise.

So far, we found the best strategy for Eve to guess Alice's phase. Let us now consider how Eve can utilize her partial information to mimic Alice in the best possible way and to "replenish" the state transmitted to Bob, based on her guess. Note that Eve's measurement already affected the transmitted state towards Bob, indicating that the best option for Eve is not to simply "amplify" the remainder beam (say by a laser amplifier of appropriate gain) and send it to Bob. To understand what is the optimal "replenish" strategy for Eve, let us consider separately each of the possible measurement outcomes for her:
\begin{enumerate}
    \item \textit{P(1,1) - } Eve measured a single photon in both the signal and the idler, which indicates that she detected both photons and the remaining quantum state in Bob's channel  is now simply vacuum
\begin{equation}
    \left|\psi\right>_{T} = \left|0\right>_{Bob}
\end{equation}
with no information of Alice's basis. To better cover her tracks, Eve can generate a new photon pair at her guessed phase and send it to Bob. Note that although Eve measured both photons, she did not necessarily measure them in the correct phase basis, which indicates that she may still introduce errors (as explained above and analyzed hereon). 
    \item \textit{P(0,1) + P(1,0) - } Eve measured only one photon, either in the signal or the idler, indicating that she failed to retrieve Alice's encoding. Consequently, Eve knows that the remaining quantum state at Bob's side is the other photon, i.e. 
\begin{equation}
    \left|\psi\right>_{T} = 
    \begin{cases}
        \left|1_{\omega}, 0_{-\omega}\right>_{Bob} \\
        \left|0_{\omega}, 1_{-\omega}\right>_{Bob}
    \end{cases}.
\end{equation}
Here, Bob will receive an erroneous state that will allow him to identify Eve's presence. It is therefore better for Eve to replace the state with a new quantum state with a random guess for Alice's phase, similar to an intercept-resend attack. However, since Eve does not know the correct basis to generate the new state, she cannot cover her tracks completely, as she will inevitably introduce errors, when her guess is wrong.
\item \textit{P(0,0) - } Eve measured no photons, indicating that the remaining quantum state becomes:
\begin{equation}
    \left|\psi\right>_{T} = \left|0\right>_{Bob} + it^{2}g_{\omega}e^{i(\phi_{A} + \phi_{E})} \left|1_{\omega}, 1_{-\omega}\right>_{Bob},
\end{equation}
which is relatively similar to an untampered state with two differences - smaller amplification during the pairs generation, $t^{2}g_{\omega}$ and additional phase, $\phi_{E}$. Here, Eve's best strategy is not to generate a new quantum state to produce her guess, but rather to manipulate the existing state in Bob's channel to obtain a closer result to the the original state, even when her guess is incorrect. Eve can achieve this by shifting the remained state by $-\phi_{E}$ to $\left|\psi\right>_{T} = \left|0\right>_{Bob} + it^{2}g_{\omega}e^{i\phi_{A}} \left|1_{\omega}, 1_{-\omega}\right>_{Bob}$ and then to parametrically amplify the state with another OPA, to obtain:
\begin{align}
    \left|\psi\right>_{T} = &\left|0\right>_{Bob} + ig_{\omega}\left(t^{2}e^{i\phi_{A}} + \alpha e^{i\phi_{A^\prime}} \right) \left|1_{\omega}, 1_{-\omega}\right>_{Bob} = \\ &\left|0\right>_{Bob} + ig_{\omega}\left(t^{2}e^{i\phi_{A}} + (1-t^2) e^{i\phi_{A^\prime}} \right) \left|1_{\omega}, 1_{-\omega}\right>_{Bob},
\end{align}
where $\phi_{A^\prime}$ is Eve's guess for Alice's phase and $\alpha\cdot g_{\omega}$ is Eve's amplification, chosen such that $\alpha = (1-t^2)$. Note that If Eve guessed correctly ($\phi_{A^\prime} = \phi_{A}$), she totally recovers Alice's original state and cannot be identified. In the limit of $t\rightarrow0$ (Eve stole the state entirely), Eve's amplification converges to a complete state generation, which is a direct generalization of the "intercept-resend attack" that we discussed before. 
\end{enumerate}

So far we discussed Eve's strategy to best cover her tracks during the steal-intercept attack. Eve's choices are summarized in table \ref{table:steal_intercept_attack_options}. Let us now analyze how Alice and Bob can still identify Eve's presence in each of the cases during the verification stage.
\begin{table}[]
    \centering
    \begin{tabular}{p{3cm}|p{5cm}p{5cm}}
         &   $ Basis_{A} = Basis_{E}$&$Basis_{A} \neq Basis_{E}$\\
         \hline
         Eve Measures (0,0)& 
     Manipulate the state into a guessed phase in the correct basis&Manipulate the state in an incorrect basis\\
 Eve Measures something else& Generate a new state in the correct basis with a guessed phase&Generate a new state in an incorrect basis\\\end{tabular}
    \caption{The best concealing choices for Eve during a steal-resend attack. }
    \label{table:steal_intercept_attack_options}
\end{table}

As table @@@ shows, Eve will either manipulates the remaining state (1st row) or generates a new one (2nd row). In both cases, Eve manipulates/generates the state in her measurement basis, where she gained the most information on the original state. Whenever Eve chooses to regenerate a new state (2nd row), Bob's state (after his phase selection $\phi_B$ and his OPA) is
\begin{equation}
\left|\psi\right>_{2} = 
\left|0\right> + 
ig_{\omega} (1 + e^{i(\phi_{A^\prime} + \phi_{B})}) \left|1_{\omega}, 1_{-\omega}\right>,
\end{equation}
which indicates that the average number of photons in his measurement is
\begin{equation}
N_{\omega} = |g_{\omega}|^{2}(2 + 2\cos(\phi_{A^\prime} + \phi_{B}))  = \begin{cases}
			0 / 4|g_{\omega}|^2, & Basis_{A} = Basis_{E} \\
            2|g_{\omega}|^2, &  Basis_{A} \neq Basis_{E}
		 \end{cases}.
\end{equation}
Here if Eve measured in the correct basis, the total phase will result in either  $\phi_{A^\prime} + \phi_{B} = \phi_{A} + \phi_{B} $ or $\phi_{A^\prime} + \phi_{B} =  \phi_{A} + \phi_{B} + \pi$, depending on Eve's probability to correctly guess Alice's phase. However in the incorrect basis, the total phase will be $\phi_{A^\prime} + \phi_{B} = \pm\frac{\pi}{2}$, indicating that Bob will measure $2|g_{\omega}|^{2}$ regardless of Alice's phase, indicating that 25\% (at least) of the regenerated states will lead to a detectable error for Alice and Bob.

When Eve manipulates the state (1st row), Bob's state is
\begin{equation}
\left|\psi\right>_{2} = 
\left|0\right> + ig_{\omega}\left(1 + t^{2}e^{i(\phi_{A} + \phi_{B})} + (1-t^2) e^{i(\phi_{A^\prime} + \phi_{B})} \right) \left|1_{\omega}, 1_{-\omega}\right>,
\end{equation}
and the average number of photons that Bob measures is
\begin{align}
N_{\omega} = |g_{\omega}|^{2}(&1 + t^{4} + (1-t^2)^2 + 2t^2\cos(\phi_{A} + \phi_{B}) +\\
&2(1 - t^2)\cos(\phi_{A^{\prime}} + \phi_{B}) + 2(1-t^2)t^2\cos(\phi_{A} - \phi_{A^{\prime}})).
\end{align}
Let us analyze this result in three cases: First, Eve measured in the wrong basis, second Eve measured in the correct basis but failed to recover the encoded bit, and third Eve recovered both the basis and the encoded bit. 
\begin{enumerate}
    \item \textit{Eve used the wrong basis:} In this case $\phi_{A} - \phi_{A^{\prime}}, \phi_{A^{\prime}} + \phi_{B} = \pm \frac{\pi}{2}$, resulting in
\begin{equation}
N_{\omega}=\begin{cases}
			2(1+T^2)|g_{\omega}|^2, & \phi_{A} + \phi_{B} = 0 \\
            2R^2|g_{\omega}|^2, &  \phi_{A} + \phi_{B} = \pi
		 \end{cases},
\end{equation} 
which will inevitably lead to detectable errors, unless $T=1$, i.e. Eve stole nothing. 
\item \textit{Eve measures in the correct basis but fails to recover the bit: } In this case $\phi_{A} - \phi_{A^{\prime}} = \pi$ and $\phi_{A^{\prime}} + \phi_{B} =  \pi + \phi_{A} + \phi_{B}$, leading to an average number of photons of
\begin{equation}
N_{\omega} =\begin{cases}
			4T^{2}|g_{\omega}|^2, & \phi_{A} + \phi_{B} = 0 \\
            4R^{2}|g_{\omega}|^2, &  \phi_{A} + \phi_{B} = \pi
		 \end{cases},
\end{equation}
which again inevitably leads to detectable errors unless $T=1$.
\item \textit{Eve guesses correctly both the basis and the bit:} In this case $\phi_{A} - \phi_{A^{\prime}} = 0$ and $\phi_{A^{\prime}} + \phi_{B} =  \phi_{A} + \phi_{B}$, resulting in 
\begin{equation}
N_{\omega} = |g_{\omega}|^{2}(2 + 2\cos(\phi_{A} + \phi_{B})),
\end{equation}
which is identical to an untampered communication, indicating that Eve succeeded to cover herself.
\end{enumerate}

\begin{figure}[ht!]
\centering\includegraphics[width=\textwidth]{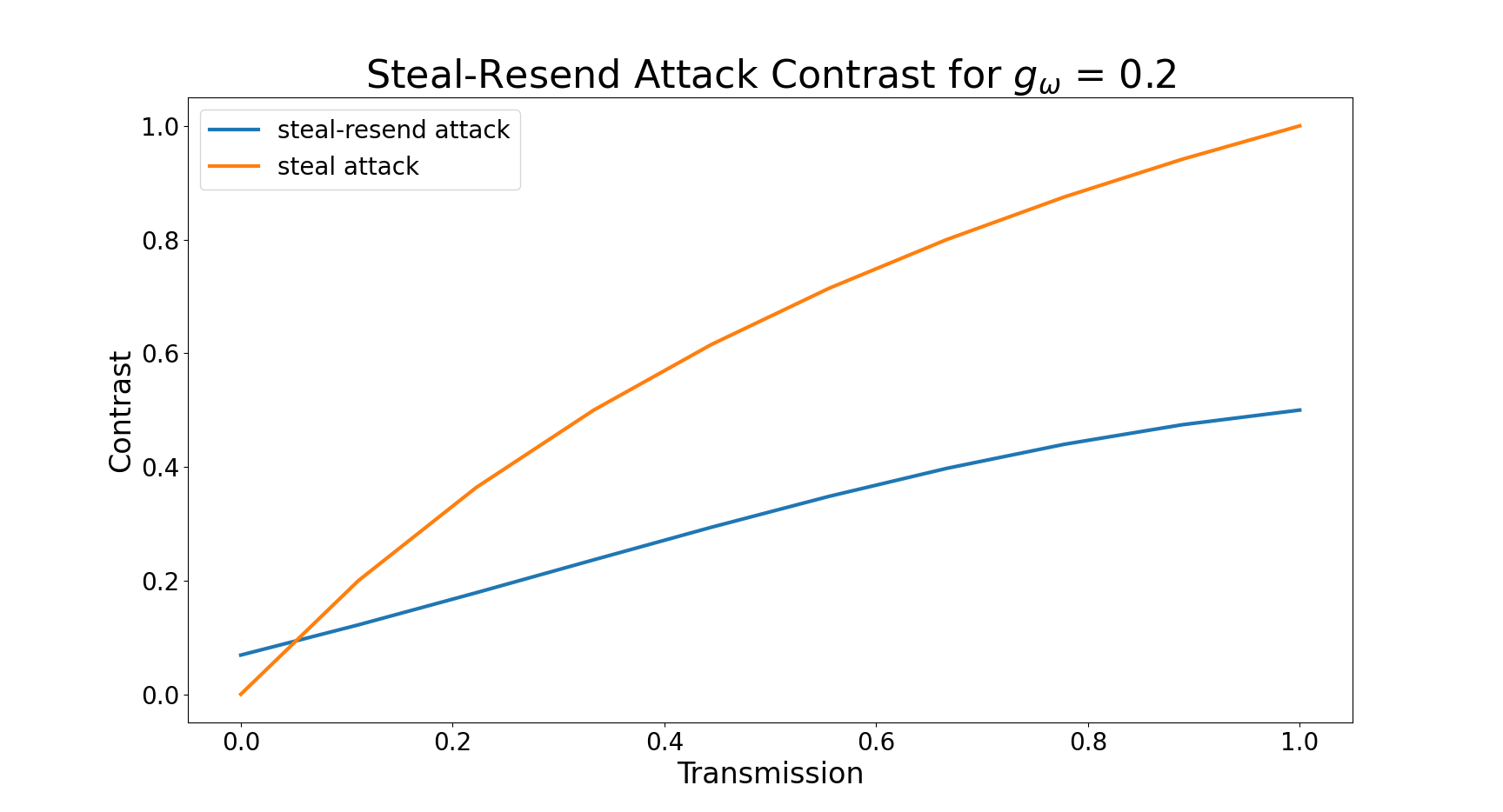}
\caption{The total contrast of steal-resend attack for amplification $g_\omega = 0.2$, compared to the contrast of steal attack. }
\label{fig:steal_resend_contranst}
\end{figure}

We analyzed above the average number of photons $N_{\omega}$ that Bob will measure in every case of Eve's attack. Using the probabilities of table \ref{table:photon_measurement_ratios} above, we can calculate 
$\left<N_\omega\right>_{\phi_A + \phi_B = 0}$ and $\left<N_\omega\right>_{\phi_A + \phi_B = \pi}$ , which will allow us to calculate both the probability of communication error and the measured interference contrast:
\begin{equation}
    V  = \frac{\left<N_\omega\right>_{\phi_A + \phi_B = 0} - \left<N_\omega\right>_{\phi_A + \phi_B = \pi}}{\left<N_\omega\right>_{\phi_A + \phi_B = 0}  + \left<N_\omega\right>_{\phi_A + \phi_B = \pi}},
\end{equation}
which remains a good witness to detect an attack (similar to the simple steal-attack).  Since an analytic expression of the contrast $V$ is complicated and tedious, we prefer to examine it graphically, as shown in figure \ref{fig:steal_resend_contranst}, where the steal-resend contrast (blue) is compared  contrast of the simple steal attack (orange). As can be seen, only near $T = 0$, where Eve steals nearly all the light, the attempts to fix the quantum state help Eve to slightly improve the contrast and to hide a bit better compared to the steal attack. However, as the transmission increases,  Eve's attempts to fix the state "turn against her" and reduce the contrast much further than a simple steal attack. This is due to the fact that the errors introduced to Alice's state recognition become too large. For Eve, it is clear that at reasonably high transmission values the steal-resend attack "defeats its purpose", and Eve will be better off to steal only a small fraction of the data, either by using a weak steal attack (with high transmission, where she receives only a small fraction of the light) or with an extreme steal-resend attack, where Eve steals the entire state, but only once in a while, hoping to remain unnoticed within the existing errors of a realistic transmission line. Nevertheless, figure \ref{fig:steal_resend_contranst}  shows clearly that the contrast is always a good witness for Alice and Bob to detect Eve, especially during a steal-resend attack, since it is \textit{always} reduced well below the contrast of the undisturbed communication (1 in an ideal setup).

% \begin{table}[]
%     \centering
%     \begin{tabular}{c|llc}
%          case&    probability&constructive&destructive\\
%          wrong basis + generation& 
%       $\frac{0.5\left(T - \frac{1}{ g_{\omega}}\right)^2}{\left(T - \frac{1}{ g_{\omega}}\right)^2+2RT + R^2+1 }$&$2|g_{\omega}|^{2}$&$2|g_{\omega}|^{2}$\\
%  wrong basis + manipulation&  $\frac{0.5(1 + 2RT + R^2)}{\left(T - \frac{1}{ g_{\omega}}\right)^2 +2RT + R^2+1 }$&2$|g_{\omega}|^{2}(1 + T^2)$&2$|g_{\omega}|^{2}(1 - T)^2$\\
%  correct guess + generation&  $\frac{0.25(R+1)^2}{\left(T - \frac{1}{ g_{\omega}}\right)^2+2RT + (R+1)^2} + \frac{0.5RT}{\left(T - \frac{1}{ g_{\omega}}\right)^2+2RT + (R-1)^2}$&$4|g_{\omega}|^{2}$&0\\
%  correct guess + manipulation&  $\frac{0.25\left(T - \frac{1}{ g_{\omega}}\right)^2}{\left(T - \frac{1}{ g_{\omega}}\right)^2+2RT + (R-1)^2}$&$4|g_{\omega}|^{2}$&0\\
%  incorrect guess + generation&  $\frac{0.25(R-1)^2}{\left(T - \frac{1}{ g_{\omega}}\right)^2+2RT + (R-1)^2} + \frac{0.5RT}{\left(T - \frac{1}{ g_{\omega}}\right)^2+2RT + (R+1)^2}$&0&$4|g_{\omega}|^{2}$\\
%  incorrect guess + manipulation&  $\frac{0.25\left(T - \frac{1}{ g_{\omega}}\right)^2}{\left(T - \frac{1}{ g_{\omega}}\right)^2+2RT + (R+1)^2}$&$4|g_{\omega}|^{2}T^{2}$&$4|g_{\omega}|^{2}(1 - T)^2$\\\end{tabular}
%     \caption{Caption}
%     \label{tab:my_label}
% \end{table}

\section{Channels Uncorrelation Measurement}
\label{section:channels_uncorrelation}

\begin{figure}[ht!]
\centering\includegraphics[width=\textwidth]{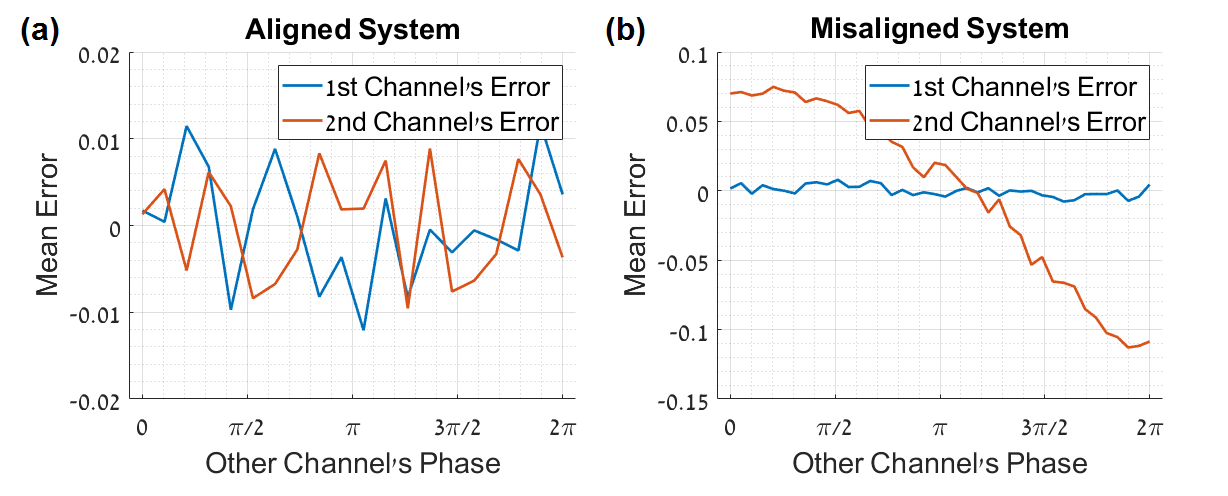}
\caption{Cross-talk evaluation: The mean error of one channel $Err_{1}(\varphi_{2}) = \left<I_{1}(\varphi_{1}, \varphi_{2}) - \right<I_{1}(\varphi_{1}, \varphi_{2})\left>_{\varphi_{2}}\right>_{\varphi_{1}}$ is recorded vs. the phase of its neighbour channel for two cases: \textbf{(a)} well aligned channels, where no correlation is observed between the channels down to the measurement noise-floor. \textbf{(b)} Misaligned spectral channels, which cause evident dependency between the phase of channel 1 and the measurement of channel 2. \textbf{(b)} also shows that correlation errors between neighbour channels are not necessarily symmetrical, depending on the leaks between the channels.}
\label{fig:uncorrelation}
\end{figure}

In our experiment of multiplexed QKD (described in section 4 of the main text), we used a $\sim100$nm bandwidth of SPDC to encode and decode 23 QKD channels simultaneously. Those frequency channels were separated and controlled using a pulse shaper, composed of a grating (to spread the spectrum into different angles), a lens, and an SLM to encode the phase (per channel). A spectrometer was used to measure all the channels simultaneously, composed of a grating, a lens and a line-CCD camera. In theory, the only mechanism for crosstalk may be the finite frequency resolution of the shaper and the spectrometer (due to the diffraction limit). In practice additional technical imperfections in both the SLM and the camera can lead to crosstalk, such as voltage leakage between neighboring pixels on the SLM. Such problems may cause unwanted correlation between neighboring channels, which lead to security problems. To rule out crosstalk between neighboring channels of the experiment (and so ensure the security of our scheme), we measured the correlation between the channels, as shown in figure \ref{fig:uncorrelation}. In optimal conditions of alignment (diffraction-limited resolution for Alice's shaper and for Bob's measurement spectrometer (CCD pixels), with perfect correspondence between them) no correlation was observed between neighboring channels down to the noise floor of our measurement (figure \ref{fig:uncorrelation}a). However, when misalignment was introduced, correlation between neighboring channels appeared in the form of leakage of the phase modulation from one channel into the measurement of the neighboring channel (see figure \ref{fig:uncorrelation}b). This correlation is an example of information leakage that may help an eavesdropper to reveal data on one channel from the measurement of its neighbor.

In addition to security verification, the uncorrelation measurement serves another purpose - to maximize the number of channels for a given experimental configuration. We used this measurement to find the smallest spectral separation between channels that preserves the uncorrelation property, allowing to maximize the number of channels within the available spectrum. Thus, the uncorrelation measurement is an important step in the calibration of our experiment. This resulted in \(~130\mu m\) wide channels with \(~30 \mu m\) wide gaps between them.

\section{Loss Detection}
\label{section:loss_detection}

To verify the security of our multiplexed QKD scheme, we demonstrated the detection of eavesdroppers. Experimentally, we identified the steal attack and quantified its detectability. As we have seen in the security analysis above (section \ref{section:steal_attack}), a steal attack reduces the average number of photons in mode $\omega$ to $N_{\omega} = \left|g_{\omega}\right|^{2}(1+T+2T\cos(\phi_{\omega}))$ (equation 4, main text) which we can measure. However, as we have seen in \ref{section:steal_attack}), the better detection parameter is the contrast (visibility) of the interference after the second crystal, which is given by
$V \equiv \frac{I_{max} - I_{min}}{I_{max} + I_{min}} = \frac{2T}{1+T}.$ (equation 4, main text). The contrast is a good witness for eavesdropping since Eve must steal some of the photons, i.e. reduce the transmission ($T$), which will lower the contrast and therefore introduce errors. 

\begin{figure}[ht!]
\centering\includegraphics[width=\textwidth]{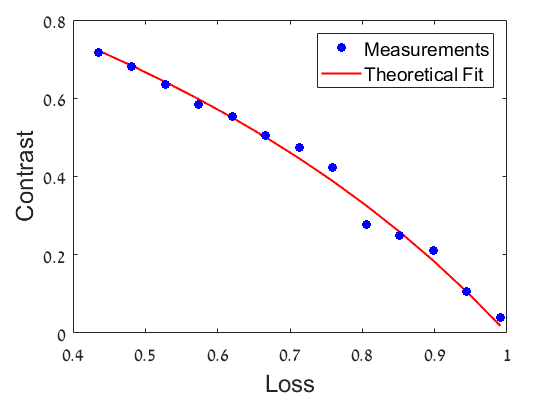}
\caption{Detection of steal-attack from measuring the contrast: The figure shows the measured contrast as function of the total transmission loss, which was varied using the SLM (amplitude modulation). Blue crosses show the measured contrast, and the red fit is calculated from equation 4 (main text).}
\label{fig:loss}
\end{figure}

Figure \ref{fig:loss} shows the measured dependence of the contrast on the loss (equation 4, main text) along with the theoretical curve, with very good agreement, indicating our ability to detect the loss from the contrast. The loss was introduced by amplitude modulation with the SLM. The only fit parameter of the theoretical curve in figure \ref{fig:loss} was the initial loss of the channel which turned out to be around $44\%$. This high loss was probably caused by a combination of the finite diffraction efficiency of the gratings and the SLM, imperfect AR-coating on the optical components, uneven propagation of the pump, the signal and the idler and other practical factors, which can all be improved in future experiments. And yet, we can clearly detect even a small additional loss of $5\%$ since it reduces the contrast visibly (compared to the measurement error). Thus, Eve will be easily detected even with realistic, rather high initial losses. Note that for Eve to obtain useful information, she must steal a substantial fraction of the light since high loss directly affects the amount of data received during the communication. 

\section{Experimental considerations on the Number of Channels}
\label{section:number_of_channels_optimization}

Although the spectrum of our SPDC source is continuous, the effective number of channels is limited by the spectral resolutions of both the shaper and the spectrometer. These spectral resolutions are governed digitally by the available number of pixels on the SLM (or linear CCD camera) within the spectrum and by the analog spectral resolution of the shaper (spectrometer), as dictated by the grating line-density and by the diffraction limit of the lenses. In this section we explain how all these parameters were tuned to optimize the number of channels in our specific experiment, laying out the "points for future improvement" that can serve future experiments.

Let us first consider the spatial width of each channel on the SLM (or spectrometer). In order to make the most out of the available pixels we optimize the diffraction limit of the lens $d \approx 2.44\frac{\lambda f}{D}$ to match approximately the pixel size. This sets an appropriate focal length for the lenses: 

\begin{equation}
f_{lens} \approx 0.4\frac{d_{pixel}D}{\lambda} 
\end{equation}
Where $d$ is the diameter of the beam at the focal point, $\lambda$ is the channel wavelength ($1560nm$ in our setup), $D$ is the beam's diameter on the lens and $f$ is the lens focal length.

The second factor to consider is the SPDC spectrum, which is limited by the phase matching bandwidth of the PPLN OPA. To utilize the most of the available SPDC, one needs to span the SPDC spectrum across the entire SLM (camera). For this we need to consider the angular dispersion of the grating on the SPDC spectrum of width $2\omega$. The diffraction angle $\theta$ of the grating adheres to
\begin{equation}
d \sin \theta = \lambda.
\end{equation}
To first order approximation $\sin \theta \approx \theta$ we obtain
\begin{equation}
d \cdot \theta_{\omega} = \frac{2\pi c}{\omega},
\end{equation}
which allows to calculate the angular span of the spectrum as
\begin{equation}
\label{equation:grating_angular_span}
d \cdot \Delta \theta = \frac{2\pi c}{\frac{\omega_{p}}{2} - \omega} - \frac{2\pi c}{\frac{\omega_{p}}{2} + \omega} = \frac{4\omega\pi c}{\frac{\omega_{p}^{2}}{4} - \omega^{2}}.
\end{equation}
Finally, we choose the grating period $d$, such that the spectrum will cover the spatial span $L$ of the SLM (camera) at the Fourier plane of the lens, which requires that
\begin{equation}
f_{lens} \cdot \Delta \theta = L.
\end{equation}
Clearly, the aperture of the focusing lenses $D$ must be sufficient to capture the complete angular spectrum of the light ($D>L$). We can now use equation \eqref{equation:grating_angular_span} to obtain the optimal grating period:
\begin{equation}
d = \frac{f_{lens}}{L}\frac{4\omega\pi c}{\frac{\omega_{p}^{2}}{4} - \omega^{2}}
\end{equation}

The final factor to consider is chromatic dispersion. The optical elements in the beam, such as lenses, filters, polarizers, etc. incur chromatic dispersion on the SPDC spectrum due to the variance of the index of refraction with frequency. Specifically, variation of the phase-sum of the signal-idler pair due to dispersion will shift the overall phase of that pair (channel), resulting in a shift of the interference pattern at this channel. Dispersion therefore should be either compensated or calibrated in order to correctly detect the transmitted information. Dispersion correction can be included in two ways: First, for relatively weak dispersion, where the phase variation across the bandwidth of a single channel is negligible, we can use the phase modulation of the SLM to pre-compensate the dispersion of each channel, which for our experiment was sufficient. However, for cases of high dispersion, as may be the case after passage through a long optical fiber, the phase variation across a single channel may no longer be small, which will lead to reduction of the interference contrast, or even to complete wash-out of the spectral interference fringes. In such a case, physical compensation of the dispersion will be necessary (e.g. with a negative-dispersion fiber or with a prism-pair, etc.).